\documentclass{article}

\usepackage{arxiv}

\usepackage[utf8]{inputenc} 
\usepackage[T1]{fontenc}    
\usepackage{hyperref}       
\usepackage{url}            
\usepackage{booktabs}       
\usepackage{amsfonts}       
\usepackage{nicefrac}       
\usepackage{microtype}      
\usepackage{placeins}
\usepackage{graphicx}
\usepackage{graphics}
\usepackage{float}
\graphicspath{{./images/}}
\title{BitMex Funding rate Correlation with Bitcoin Exchange rate}

\author{
  Sai Srikar Nimmagadda\thanks{Sai Srikar Nimmagadda and Ammanamanchi Pawan Sasanka are recent graduates of Manipal Institute of Technology.} \\
  Department of Computer Science and Engineering\\
  Manipal Institute of Technology\\
  Manipal, 576104 \\
  \texttt{sai.srikar@learner.manipal.edu} \\
   \And
  Ammanamanchi Pawan Sasanka\thanks{}\\
  Department of Information and Communication Technology\\
  Manipal Institute of Technology\\
  Manipal, 576104 \\
  \texttt{pawan.sasanka@learner.manipal.edu} \\
}

\begin{document}
\maketitle

\begin{abstract}
This paper examines the relationship between Inverse Perpetual Swap contracts, a Bitcoin derivative akin to futures and the margin funding interest rates levied on BitMex exchange. This paper proves the Heteroskedastic nature of funding rates and goes onto establish a causal relationship between the funding rates and the Bitcoin inverse Perpetual swap contracts based on Granger causality. The paper further dwells into developing a predictive model for funding rates using best-fitted GARCH models. Implications of the results are presented, and funding rates as a predictive tool for gauging the market trend is discussed. 
\end{abstract}


\section{Introduction}
\noindent Bitcoin, a brainchild of Satoshi Nakamoto, is the first cryptocurrency to be created based on the blockchain technology. The rise of Bitcoin has given birth to several alternative cryptocurrencies, and yet, Bitcoin still boasts of the highest market capitalization and is the leading cryptocurrency in the digital asset space. Bitcoin has outperformed several assets and asset classes, and due to increasing speculation, it is now gradually becoming one of the highest traded commodities. Increasing institutional involvement prompted by the classification of Bitcoin as a commodity by SEC  has become a proponent of this trend. The inflection of Bitcoin as a speculative asset has brought in a plethora of trading instruments with inverse Perpetual swaps offered by BitMex being the most traded Bitcoin derivatives in the market. The annual volume traded on this instrument on BitMex(Bitcoin Mercantile Exchange) has reached levels beyond 1 Trillion USD. It is one of the reasons for BitMex to be of high relevance in this paper. BitMex provides margin trading of a plethora of contracts, some of which are standard futures contracts with settlement date, and one with Perpetual contracts with no settlement date or expiry. 

There are several other Bitcoin derivative products offered by Bitfinex, Deribit, and the more institutional CME and CBOE. However, none of the aforementioned match the trading volume generated by BitMex. There have been efforts to understand the relationship between spot markets and the derivatives market, specifically the Futures market of Bitcoin by \cite{doi:10.1002/fut.22004} and \cite{AKYILDIRIM2019}. Furthermore, there has been noteworthy literature in studying the relationship between margin lending and its implications on equities by \cite{RePEc:fip:fedbne:y:2003:p:31-50}. However, this relationship is seldom explored in Bitcoin and Cryptocurrency markets. Several papers are published aimed at further understanding the relationship of Bitcoin with factors such as volatility by \cite{DYHRBERG201685} and \cite{RePEc:eee:ecolet:v:158:y:2017:i:c:p:3-6}, On-Return volatility by \cite{BOURI2017}, liquidity by \cite{RePEc:eee:ecolet:v:165:y:2018:i:c:p:58-61}, market inefficiency by \cite{RePEc:eee:ecolet:v:168:y:2018:i:c:p:21-24} and transactions by \cite{Greaves2015UsingTB}. There is limited research examining the relationship between an exchange and the price of Bitcoin as well. To further the literature of Bitcoin in finance, this study focuses on a novel aspect where the correlation of Bitcoin price with margin interest rates levied on BitMex Perpetual Swap Contracts is studied. The correlation of interest rates in liquidity which leads to price fluctuations (volatility) is studied by \cite{Brunnermeier01062009}. The influence of interest rates on the liquidity of BitMex derivatives market is beyond the scope of this paper. The current study in the paper could be of significant interest to market stakeholders and investors to improve their investment decisions. This paper deals with predominantly three aspects :

\begin{itemize}
\item To prove BitMex funding rates are Heteroskedastic in nature. 
\item To correlate the funding rates with market price of BTC/USD Perpetual swap contracts.
\item Implication of these results and funding rates as a tool to predict market trends. 
\end{itemize}

This paper is organized as follows Section~\ref{section:mex} gives an introduction to BitMex Perpetual contracts and how funding works in the exchange. In Section~\ref{section:DM}, a summary of the data used, followed by a comprehensive discussion about the methodology is presented. Section~\ref{section:DM} also has sub-sections that elaborate on the tests implemented. Section~\ref{section:results} comprises of the results obtained. Lastly, the conclusions are presented in Section~\ref{section:conclusion}.

\section{BitMex, Perpetual Contracts and Funding} \label{section:mex}
The Perpetual swap derivatives, unlike futures, are Perpetual and have no settlement date. Contrary to the Futures Contract, the Perpetual Contracts trade close to the Index price. Here the Index price is used as a reference. Such margin structure causes these contracts to not trade at markedly higher prices than the base price. Here, Funding acts as a way to anchor the Contracts to the spot-price. It acts as a balancing force. The Funding comprises of three 8 hour periods where payments are settled between the buyer and seller. A negative Funding Rate implies that short positions pay long positions and vice versa if the rate is positive.
Furthermore, the Funding Payout is made only if the traders are holding a position during the Funding Timestamp. The 8 hourly Funding Timestamps are 4:00 UTC,12:00 UTC and 8:00 UTC. For example, if a trader takes a long position when the Funding rate is positive and holds the position until either of the payout Timestamps, the trader pays this Funding to the short positioned traders. The Funding according to \cite{BitMex} is calculated as follows
\begin{equation}
   Funding = Position Value * FundingRate(F)
\end{equation}
The Funding Rate is calculated based on the interest rate and Premium/Discount. The Funding rate acts as the interest rate to be paid between buyers and sellers akin to margin-trading markets. The Interest rate is the interest rates between the Base currency and Quote currency. The Interest Rate according to \cite{BitMex} is defined as \vspace{3mm}
\begin{equation}
    Interest Rate(I) ={\frac{Interest Quote Index - Interest Base Index}{Funding Interval(FI)}}
    \vspace{3mm}
\end{equation}
where,\\
Interest Base Index = The Interest Rate for borrowing the Base currency
\newline
    Interest Quote Index = The Interest Rate for borrowing the Quote currency
\newline
    Funding Interval = 3, because funding occurs three times a day.  

Further, the Premium/discount component can be comprehended as Premium or discount to the Mark Price. The Premium Index, $P$,  will be used to equalize the next Funding Rate to levels consistent to spot-price.
\newline
\newline The Premium, according to \cite{BitMex}, is calculated as follows : \vspace{2mm}
\begin{equation}
   Premium Index(P) = {\frac{(Max(0, IBP - Mark Price) - Max(0, Mark Price - IAP))}{Spot Price + Fair Basis}}
\end{equation}
where, 
\newline Impact Bid Price(IBP) = The average fill price to execute the Impact Margin Notional on the Bid side
\newline Impact Ask Price(IAP) = The average fill price to execute the Impact Margin Notional on the Ask side
\vspace{2mm}
\begin{equation}
Fair Basis  = Index Price * (1 + Funding Basis)
\end{equation}
And 
\begin{equation}
Funding Basis = F * {\frac{Time Until Funding}{FI}}
\end{equation}
The Impact Margin Notional is used to determine the impact of 0.1 BTC Notional on the depth of the order book and is used to measure Impact Bid or Impact Ask Price.

Every minute, BitMex calculates Premium/Discount (P) component and Interest Component (I). The 8-hour TWAP of Premium/Discount component and the Interest component is calculated over the minute time series. Finally, a $\pm0.05\%$  dampener is added to 8-hour components of Premium/Discount and Interest. The Funding Rate is then calculated as
\begin{equation}
    F = P + clamp(I -P, 0.05\%, -0.05\%)
\end{equation}
This Funding Rate is then used to determine the Funding amount to be paid or received by the trader over their positions at Funding Timestamp. It should be noted that BitMex caps the Absolute Funding Rates at not more than 75\% of Initial Margin - Maintenance margin, furthermore, the Funding Rate can not exceed more than 75\% of the maintenance margin.

\section{Data and Methodology}\label{section:DM}
\subsection{Data}
This paper uses 8-hour time series of Bitcoin price, unlike the standard daily or hourly, from Bitstamp, the largest exchange \cite{RePEc:eee:intfin:v:36:y:2015:i:c:p:18-35}, as it is one of the exchanges used in the formulation of BitMex Composite Index. Its high availability and long record of exchange rates is also one of the reasons. The use of 8-hour time series facilitates smoother proof of correlation with Funding Rate, which also follows 8 hour periods. Along with the exchange rate the 8-hour time series of Funding Rate from BitMex is also used. The two time-series data span over the same periods - From June 4th, 2016  to Oct 3rd, 2019 (3649 observations). The Figure 1 and Figure 2 plot the Funding rate, the Log Funding Rate respectively and Figure 3 plots the price of Bitcoin during the period of study. The log-returns of funding are calculated as 
\begin{equation}
        Funding (F) = Ln(F_{t})- Ln(F_{t-1})
\end{equation}
\begin{figure}[H]
    \centering
    \frame{\includegraphics[scale = 0.4]{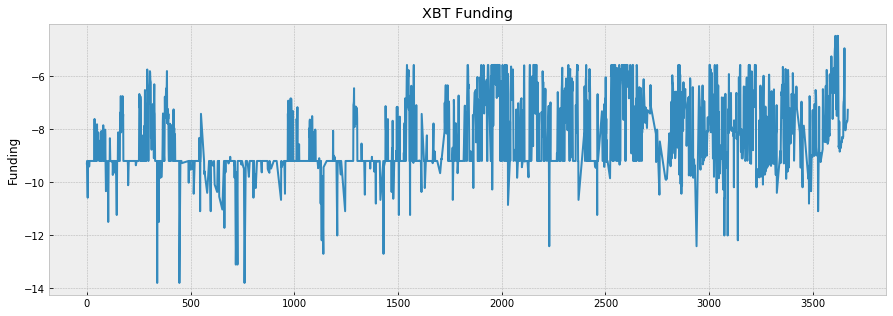}}
    \caption{Logarithmic Funding rate}
    \label{fig:my_label1}
\end{figure}
\begin{figure}[H]
    \centering
    \frame{\includegraphics[scale=0.4]{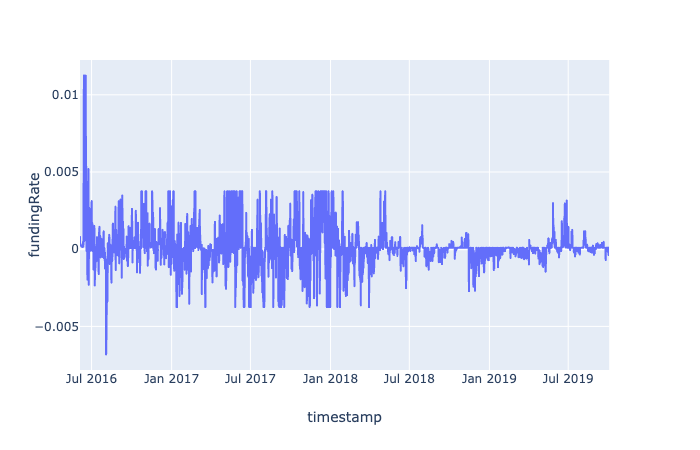}}
    \caption{Funding Rate}
    \label{fig:my_label2}
\end{figure}
\begin{figure}[h!]
    \centering
    \frame{\includegraphics[scale = 0.4]{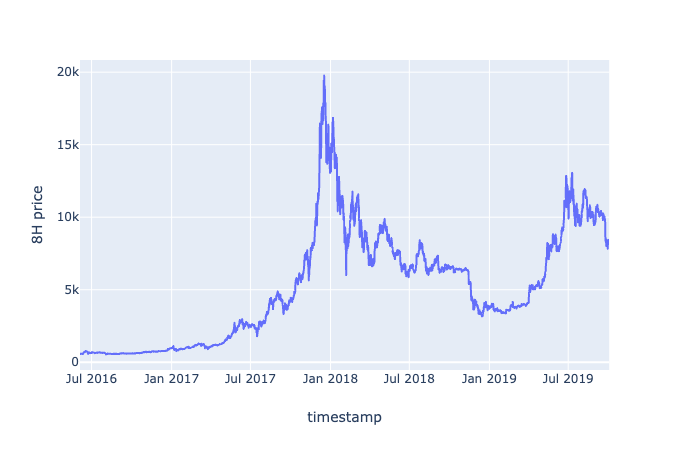}}
    \caption{BTC Price against time}
    \label{fig:my_label}
\end{figure}
\subsection{Methodology}
Funding rate functions as Interest rates for margin lending, see Section~\ref{section:mex}. The methodology to model Funding Rate as Heteroskedastic is adopted from the alternative hypothesis for Engle’s  ARCH Test for heteroskedasticity\cite{RePEc:ecm:emetrp:v:50:y:1982:i:4:p:987-1007}. Previous literature in modelling interest rates in the US\cite{RePEc:bla:jfinan:v:38:y:1983:i:2:p:635-46}, \cite{10.2307/2328983} have shown no clear evidence of Conditional heteroskedasticity. Similarly, \cite{RePEc:ecm:emetrp:v:50:y:1982:i:4:p:987-1007} ARCH test for heteroskedasticity in the residuals of US and UK real interest rates show no evidence either. However, there is evidence of this process in interest rates in the UK\cite{RePEc:eee:jbfina:v:18:y:1994:i:1:p:153-165}. Therefore, it is essential to test for heteroskedasticity due to the varying nature of Interest rates. A positive test for heteroskedasticity allows modelling of Funding rate under GARCH process viable. However, after this, actual correlation causation between Funding Rate and the price of Bitcoin needs to be established to support the aim of the paper. It is done by first testing for stationarity of the two series via ADF test, a form of unit root testing, followed by testing for Granger’s Causality test to prove Funding Rate Granger Causes Price.
\subsubsection{Heteroskedasticity test}
The alternative test for Heteroskedastic process follows test for heteroskedasticity \cite{RePEc:ecm:emetrp:v:50:y:1982:i:4:p:987-1007} where an uncorrelated time series can show dependency caused by a conditional variance process. The standard ARCH(1) process is defined by 

\begin{equation}
    		r_{t} = \sigma_{t}w_{t}
\end{equation}
\begin{equation}
    w_{t} = white\:noise
\end{equation}
\begin{equation}
    \sigma_{t} = {\sqrt{\omega+\alpha_{1}r^2_{t-1}}}
\end{equation}
Here, $r_{t}$ is defined as returns and $e_{t}$ has 0 mean and variance of 1. Let the Funding rate time series be defined as
\begin{equation}
    y_{t} = u_{t} + \epsilon_{t}
\end{equation}
$u_{t}$ = conditional mean
\newline $\epsilon_{t}$ = innovation process with mean 0 
the residual series is defined by 
\begin{equation}
    e_{t} = y_{t} - \hat\mu_{t}
\end{equation}
This alternative hypothesis test is auto-correlation in the squared residuals, given by the regression
\begin{equation}
    H_{a} = u^2_{t}=\alpha_{0} +\alpha_{1}u^2_{t-1}+\alpha_{2}u^2_{t-2}...+\alpha_{3}u^2_{t-p} + v_{t}
\end{equation}
Here $u_{t}$ is white noise error process,The null hypothesis is given by
\begin{equation}
    H_{a} = \alpha_{1} = \alpha_{2} = ... = \alpha_{m} = 0
\end{equation}
To conduct the alternative hypothesis, a lag of n is used, where n is determined by tests such as AIC \cite{1100705}, where 
\begin{equation}
    AIC = 2k - 2Log(\hat\theta)
\end{equation}
Here k is number of estimated parameters and $\hat\theta$ cap as maximum log-likelihood. The Funding Rate is normalized as
\begin{equation}
    FundingRate = FundingRate_{t} - FundingRate_{t-1}
\end{equation}
The alternative ARCH test results are presented in Table~\ref{tab:Table1}
\begin{table}[h!]
\centering
\caption{ARCH Test results}
\begin{tabular}{  c  c  c  } 
\toprule
Probability of F & 0.000 \\ 
\hline
Probability of LM Statistic (Chi-Square(1)) & 0.000 \\ 
\hline
\multicolumn{2}{c}{Variables} \\
\hline
Probability of C & 0.000 \\
\hline
value of alpha1(squared residuals) & 0.284  \\ 
\bottomrule
\end{tabular}
\label{tab:Table1}
\end{table}
\FloatBarrier
Under the null hypothesis \begin{equation}
    \alpha_{1}!= 0
\end{equation}
therefore showing ARCH effects and hence proving heteroskedasticity. $\alpha_{1} = 0.284$ and is statistically significant. Since $\alpha_{1}$ is significant, $R^2$, the coefficient of determination will be relatively high, showing the presence of ARCH effects. Hence hypothesis of no ARCH effects is rejected.
On modelling the ARCH process, the time series exhibits white noise. 
\begin{figure}[H]
    \centering
    \frame{\includegraphics[scale = 0.3]{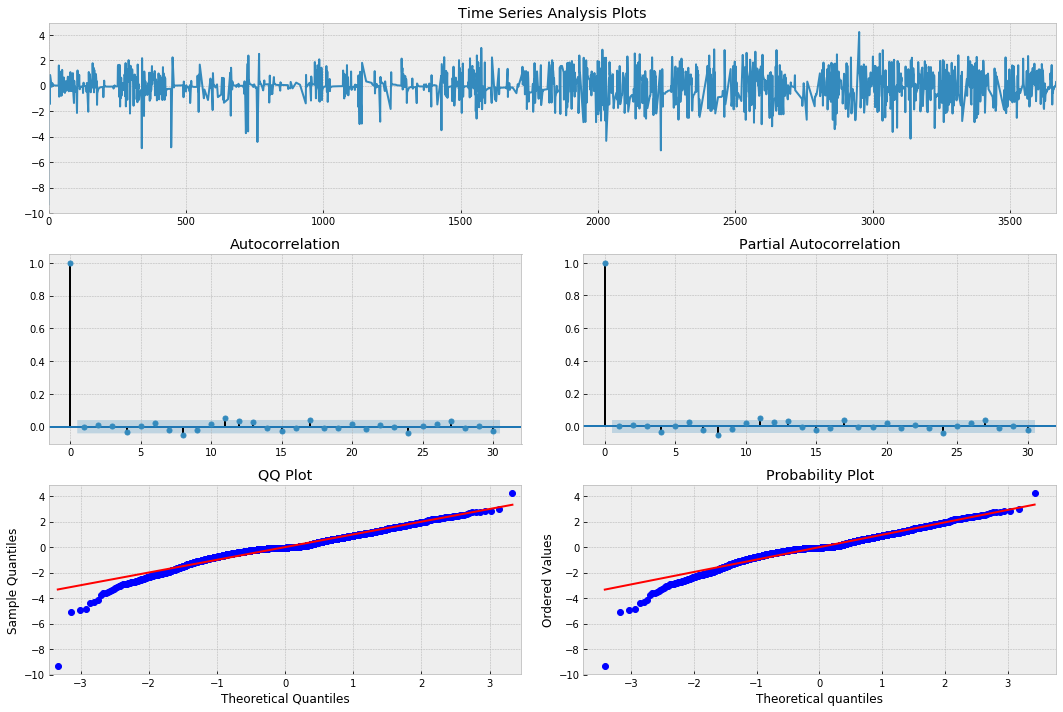}}
    \caption{Time Series Analysis plots}
    \label{fig:my_label3}
\end{figure}
However, the squared of the series shows successive decaying lags.
\begin{figure}[H]
    \centering
    \frame{\includegraphics[scale = 0.3]{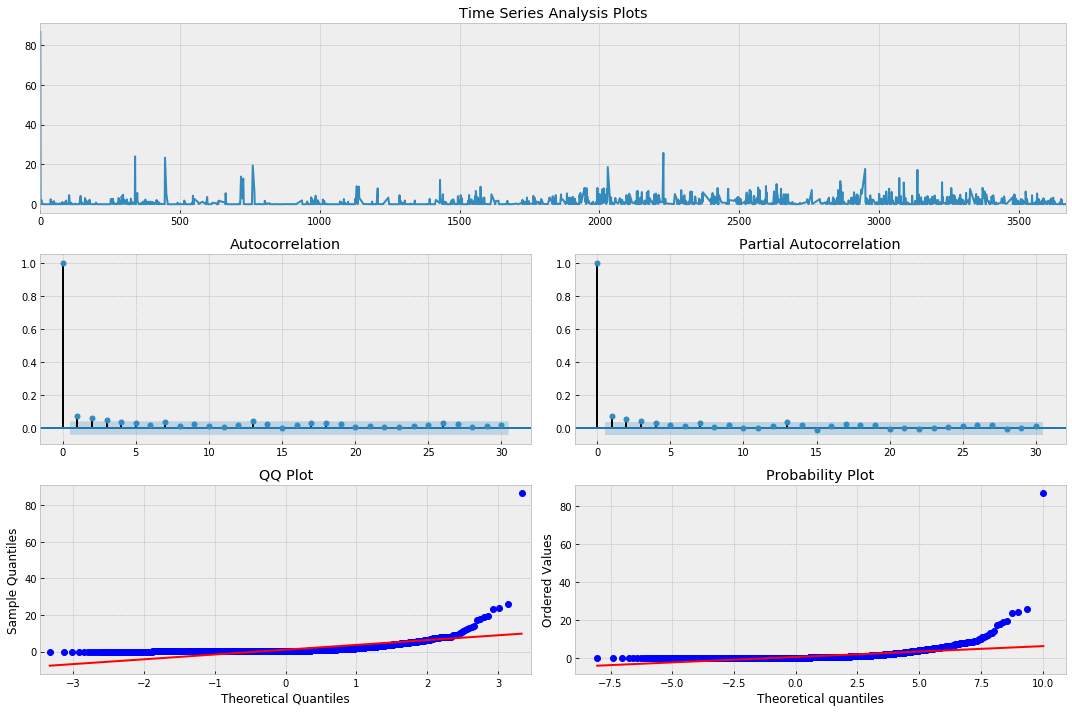}}
    \caption{Time Series Analysis plots after squared series}
    \label{fig:my_label4}
\end{figure}
The Empirical tests for conditional heteroskedasticity process are positive when squared plots of Auto Correlation(ACF) and Partial Auto Correlation (PACF) show substantial evidence via the decay of successive lags. 
\subsubsection{Unit Root Test}
To perform the Granger Cause test, the two time-series data of Funding Rate and Exchange Rate are to be tested for stationarity. If the time-series exhibits non-stationarity, it is transformed into a stationary series. ADF test \cite{10.2307/2286348} on both time-series convey the following results. The lag lengths are determined using AIC \cite{1100705}. Funding Rate is stationary at level and I(1) via ADF test of unit root testing and Exchange Rate is stationary at I(1) only. Since p-values are less than 0.05 and are significant, the null hypothesis is rejected.

\begin{table}[h!]
\centering
\caption{Stationarity in Funding Rate}
\begin{tabular}{ c  c  c  c  c  c  c } 
\toprule
Test for Unit root in & Exogenous & t-Statistic & Probability & 1\% level & 5\% level & 10\% level\\ 
 \hline\hline
 Level &  Constant & -9.176579 & 0.0000 & -3.431965 & -2.862139 & -2.567132\\ 
 \hline
Level & Constant and Linear Trend & -9.345478 & 0.0000 & -3.960568 & -3.411044 &  -3127399\\
 \hline
 Level & None & -9.010413 & 0.0000 & -2.56599 & -1.94011 & -1.616641\\
 \hline
 First difference & Constant & -16.35916 & 0.0000 & -3.431971 & -2.86141 & -2.567134\\
\hline
First difference & Constant and Linear Trend & -16.35687 & 0.0000 & -3.960576 & -3.411048 & -3.127342\\
\hline
First difference & None & -16.6138 & 0.0000 & -2.565601 & -1.940911 & -1.616641\\
\bottomrule
\end{tabular}
\label{tab:Table2}
\end{table}
\FloatBarrier

\begin{table}[h!]
\centering
 \caption{Stationarity in Exchange Rate}
 \begin{tabular}{ c  c  c  c  c  c  c } 
 \toprule
Test for Unit root in & Exogenous & t-Statistic & Probability & 1\% level & 5\% level & 10\% level\\ 
 \hline\hline
 Level &  Constant & -1.743900 & 0.49090 & -3.431971 & -2.862141 & -2.567134\\ 
\hline
Level & Constant and Linear Trend & -2.234493 & 0.4696 & -3.960576 & -3.411048 & -3.127342\\
 \hline
 Level & None &  -1.617692 & 0.0998 & -2.565601 & -1.940911 & -1.616641\\
 \hline
 First difference & Constant & -10.16478 & 0.0000 & -3.431971 & -2.862141& -2.567134\\
\hline
First difference & Constant and Linear Trend & -10.16079 & 0.0000 & -3.960576 & -3.411048 & -3.127342\\
\hline
First difference & None & -10.14073 & 0.0000 & -2.565601 & -1.940911 & -1.616641\\
\bottomrule
\end{tabular}
\label{tab:Table3}
\end{table}
\FloatBarrier


\subsubsection{Granger Causality}
To determine the Granger Causality test\cite{RePEc:ecm:emetrp:v:37:y:1969:i:3:p:424-38}, \cite{RePEc:eee:dyncon:v:2:y:1980:i:1:p:329-352} is performed over the stationary time-series of first differenced Funding Rate and first differenced Exchange Rate. Here, again the optimum lags obtained from AIC \cite{1100705} is applied to get the best fit. 
\begin{table}[h!]
\centering
 \caption{Causality test results}
 \begin{tabular}{ c  c  c } 
 \toprule
Null Hypothesis & F-statistic & Probability\\ 
 \hline\hline
 8 Hour price does not Granger Cause FundingRate &  4.28999 & 5.E-08 \\ 
 \hline
 Funding Rate does not Granger Cause 8 hour price & 7.11106 & 1.E-15 \\
 \bottomrule
\end{tabular}
\label{tab:Table4}
\end{table}
\FloatBarrier
The P-values for both Granger Causes fall under the critical p-value implying Granger Causality exists between Funding Rate and Price. To add on the findings, the results suggest that even the exchange rate happens to Granger Cause Funding Rate. 

\section{Results}\label{section:results}
Under the null of white noise, GARCH(p,q) effects are locally equivalent alternatives to ARCH(q). This is implied from the equivalence of white noise disturbances against GARCH(p,q) disturbances in linear regression models to disturbances of ARCH(q) \cite{RePEc:eee:ecolet:v:37:y:1991:i:3:p:265-271}. The tests for ARCH effects are generally valid to GARCH effects as well, thereby validating previous tests of ARCH models to GARCH models. To conclude the selection of model, various parameters such as AIC \cite{1100705},  HQC \cite{10.2307/2985032} and SIC \cite{10.2307/2958889} under normal (Gaussian) distributions are used to get the best fit. An EGARCH(1,1) model is proven to generate the best fit to model Funding Rate. The AIC, HQC and SIC of the models are listed in Table~\ref{tab: Table5}, where the EGARCH model has the most suitable Information Criterion.
\begin{table}[h!]
\centering
\caption{Information Criterion for various models}
 \begin{tabular}{ c  c  c  c } 
 \toprule
Model & AIC & SIC & HQC\\ 
 \hline\hline
GARCH(1,1) & -11.56876 & -11.56366 & -11.57469 \\ 
 \hline
TARCH & -11.57676 & -11.56826 & -11.57373\\
 \hline
EGARCH(1,1) & -11.60566 & -11.59716 & -11.60263\\
 \hline
PARCH(1,1,1) & -11.56799 & -11.55949 & -11.57273\\
\hline
IGARCH(1,1) & -11.45785 & -11.45615 & -11.45725\\
\bottomrule
\end{tabular}
\label{tab: Table5}
\end{table}
\FloatBarrier

From the above table, the EGARCH(1,1) model is more suitable for modelling Funding Rate after evaluating AIC, SIC and HQC for their least values. An EGARCH model or an Exponential GARCH model \cite{RePEc:ecm:emetrp:v:59:y:1991:i:2:p:347-70} is defined by :
\begin{equation}
    log\sigma^2_{t} =\omega+\alpha_{1}Z^2_{t-1} + \gamma_{1}[|Z_{t-1}|- E(|Z_{t-1}|)] + \beta1og\sigma^2_{t-1}  
\end{equation}
 
for $\alpha_{1} > 0, \beta_{1} > 0, \gamma_{1} > 0$ and $\omega > 0.$
From our model the values are $\alpha_{1}= 0.1291, \beta_{1}= 0 .5466, \gamma_{1}= 0.4126 , \omega = 0.5218$
The results present a suitable model for Funding Rates, and the findings could be implemented in predicting Funding Rates, and due to causality correlation, in turn, forecast trends in the Bitcoin derivative markets.

\section{Conclusion}\label{section:conclusion}
The paper begins with a proof of heteroskedasticity of Funding Rates to correlating Funding Rates to Exchange Rate and finally, modelling the Funding rates. This could be instrumental in further forecasting trends of Bitcoin as Extremely high Funding Rates would causate high prices and vice versa due to correlation. The behaviour Funding Rates could be of greater interest in market making and Funding payouts which opens up scope to further the literature of Funding Rates and its effect on the market. The effect on Funding payouts is of vital interest as BitMex, due to its inherent Peer-to-Peer lending enables generating interest on margin lent. Future work would be to evaluate returns from a model where positions are decided by Funding rate values and then compared to the Naive Buy and Hold method.  By predicting Funding Rates, a market participant could make more informed decisions regarding their positions and future investments. This paper could be of significant interest to Financial Institutions venturing into Bitcoin derivatives market.

\bibliographystyle{apalike}  
\bibliography{template}

\begin{thebibliography}{}

\bibitem[{Akaike}, 1974]{1100705}
{Akaike}, H. (1974).
\newblock A new look at the statistical model identification.
\newblock {\em IEEE Transactions on Automatic Control}, 19(6):716--723.

\bibitem[Akyildirim et~al., 2019]{AKYILDIRIM2019}
Akyildirim, E., Corbet, S., Katsiampa, P., Kellard, N., and Sensoy, A. (2019).
\newblock The development of bitcoin futures: Exploring the interactions
  between cryptocurrency derivatives.
\newblock {\em Finance Research Letters}.

\bibitem[Baur and Dimpfl, 2019]{doi:10.1002/fut.22004}
Baur, D.~G. and Dimpfl, T. (2019).
\newblock Price discovery in bitcoin spot or futures?
\newblock {\em Journal of Futures Markets}, 39(7):803--817.

\bibitem[Bitmex, 2014]{BitMex}
Bitmex (2014).
\newblock {Bitmex Perpetual Contract Guide}.
\newblock
  \url{https://www.bitmex.com/app/perpetualContractsGuide#Funding-Rate-Calculations}.

\bibitem[Bouri et~al., 2016]{BOURI2017}
Bouri, E., Azzi, G., and Dyhrberg, A.~H. (2016).
\newblock On the return-volatility relationship in the bitcoin market around
  the price crash of 2013.

\bibitem[Brandvold et~al., 2015]{RePEc:eee:intfin:v:36:y:2015:i:c:p:18-35}
Brandvold, M., Molnár, P., Vagstad, K., and Andreas~Valstad, O.~C. (2015).
\newblock Price discovery on bitcoin exchanges.
\newblock {\em Journal of International Financial Markets, Institutions and
  Money}, 36(C):18--35.

\bibitem[Brauneis and Mestel, 2018]{RePEc:eee:ecolet:v:165:y:2018:i:c:p:58-61}
Brauneis, A. and Mestel, R. (2018).
\newblock Price discovery of cryptocurrencies: Bitcoin and beyond.
\newblock {\em Economics Letters}, 165(C):58--61.

\bibitem[Brunnermeier and Pedersen, 2009]{Brunnermeier01062009}
Brunnermeier, M.~K. and Pedersen, L.~H. (2009).
\newblock Market liquidity and funding liquidity.
\newblock {\em Review of Financial Studies}, 22:2201--2238.
\newblock Market liquidity and the funding of traders are mutually reinforcing,
  giving rise to "liquidity phenomena" like fragility, commonality and flight
  to quality.

\bibitem[Chan et~al., 1992]{10.2307/2328983}
Chan, K.~C., Karolyi, G.~A., Longstaff, F.~A., and Sanders, A.~B. (1992).
\newblock An empirical comparison of alternative models of the short-term
  interest rate.
\newblock {\em The Journal of Finance}, 47(3):1209--1227.

\bibitem[Dickey and Fuller, 1979]{10.2307/2286348}
Dickey, D.~A. and Fuller, W.~A. (1979).
\newblock Distribution of the estimators for autoregressive time series with a
  unit root.
\newblock {\em Journal of the American Statistical Association},
  74(366):427--431.

\bibitem[Dyhrberg, 2016]{DYHRBERG201685}
Dyhrberg, A.~H. (2016).
\newblock Bitcoin, gold and the dollar – a garch volatility analysis.
\newblock {\em Finance Research Letters}, 16:85 -- 92.

\bibitem[Engle, 1982]{RePEc:ecm:emetrp:v:50:y:1982:i:4:p:987-1007}
Engle, R. (1982).
\newblock Autoregressive conditional heteroscedasticity with estimates of the
  variance of united kingdom inflation.
\newblock {\em Econometrica}, 50(4):987--1007.

\bibitem[Evans et~al., 1994]{RePEc:eee:jbfina:v:18:y:1994:i:1:p:153-165}
Evans, L., Keef, S., and Okunev, J. (1994).
\newblock Modelling real interest rates.
\newblock {\em Journal of Banking and Finance}, 18(1):153--165.

\bibitem[Fortune, 2003]{RePEc:fip:fedbne:y:2003:p:31-50}
Fortune, P. (2003).
\newblock {Margin requirements across equity-related instruments: how level is
  the playing field?}
\newblock {\em New England Economic Review}, pages 31--50.

\bibitem[Granger, 1969]{RePEc:ecm:emetrp:v:37:y:1969:i:3:p:424-38}
Granger, C. (1969).
\newblock Investigating causal relations by econometric models and
  cross-spectral methods.
\newblock {\em Econometrica}, 37(3):424--38.

\bibitem[Granger, 1980]{RePEc:eee:dyncon:v:2:y:1980:i:1:p:329-352}
Granger, C. (1980).
\newblock Testing for causality: A personal viewpoint.
\newblock {\em Journal of Economic Dynamics and Control}, 2(1):329--352.

\bibitem[Greaves and Au, 2015]{Greaves2015UsingTB}
Greaves, A.~S. and Au, B. (2015).
\newblock Using the bitcoin transaction graph to predict the price of bitcoin.

\bibitem[Hannan and Quinn, 1979]{10.2307/2985032}
Hannan, E.~J. and Quinn, B.~G. (1979).
\newblock The determination of the order of an autoregression.
\newblock {\em Journal of the Royal Statistical Society. Series B
  (Methodological)}, 41(2):190--195.

\bibitem[Katsiampa, 2017]{RePEc:eee:ecolet:v:158:y:2017:i:c:p:3-6}
Katsiampa, P. (2017).
\newblock Volatility estimation for bitcoin: A comparison of garch models.
\newblock {\em Economics Letters}, 158(C):3--6.

\bibitem[Lee, 1991]{RePEc:eee:ecolet:v:37:y:1991:i:3:p:265-271}
Lee, J. H.~H. (1991).
\newblock A lagrange multiplier test for garch models.
\newblock {\em Economics Letters}, 37(3):265--271.

\bibitem[Marsh and Rosenfeld, 1983]{RePEc:bla:jfinan:v:38:y:1983:i:2:p:635-46}
Marsh, T.~A. and Rosenfeld, E.~R. (1983).
\newblock Stochastic processes for interest rates and equilibrium bond prices.
\newblock {\em Journal of Finance}, 38(2):635--46.

\bibitem[Nelson, 1991]{RePEc:ecm:emetrp:v:59:y:1991:i:2:p:347-70}
Nelson, D.~B. (1991).
\newblock Conditional heteroskedasticity in asset returns: A new approach.
\newblock {\em Econometrica}, 59(2):347--70.

\bibitem[Schwarz, 1978]{10.2307/2958889}
Schwarz, G. (1978).
\newblock Estimating the dimension of a model.
\newblock {\em The Annals of Statistics}, 6(2):461--464.

\bibitem[Wei, 2018]{RePEc:eee:ecolet:v:168:y:2018:i:c:p:21-24}
Wei, W.~C. (2018).
\newblock {Liquidity and market efficiency in cryptocurrencies}.
\newblock {\em Economics Letters}, 168(C):21--24.

\end{thebibliography}

\end{document}